# Metamaterials with index ellipsoids at arbitrary k-points

Wen-Jie Chen [1], John B. Pendry [2], and C. T. Chan [1]

[1] Department of Physics and the Institute for Advanced Study, The Hong Kong University of Science and Technology, Hong Kong, China

[2] Condensed Matter Theory Group, Physics Department, Imperial College, London SW7 2AZ, U.K.

We propose and investigate a new type of metamaterial structure composing of several interpenetrating wire meshes. Calculated band structures show that they exhibit index ellipsoids locating at nonzero k-point in long wavelength limit. We can comprehend these quasistatic modes by Poison's equation and find these modes do not rely on the detailed geometry of the wires but the connectivities of the wires. We can engineer the locations of index ellipsoid by designing the connectivities of the wire meshes.

**\*** Corresponding author: *phchan@ust.hk*

Metamaterials[1-3], as one kind of artificial subwavelength structures, have been proposed to realize exotic effective permittivity and permeability that does not exist in natural materials. Due to its exotic response to electromagnetic fields, many unique phenomena have attracted tremendous interest, such as negative refraction[4-6] and electromagnetic cloak[7-9]. In most of the previously designed metamaterials, their responses to electric and magnetic fields are tailored by the built-in resonance structures, such as wire medium[1] and split ring resonator[2], which exhibit negative permittivity and permeability for some finite frequency ranges. Thus their unnatural effective constitutive parameters do not reply on the arrangement of the resonant structures. The equifrequency surface always forms an index ellipsoid (double negative medium) or hyperboloid (hyperbolic medium) centered at the zero k-point. If the local effective permittivity and permeability are well defined, we expect a linear dispersion at low frequency in the long wave limit, and $\omega \to 0$ as $k \to 0$.

In this paper, we propose a new type of wire metamaterials[1,10-13] possessing one or more index ellipsoids positioned at nonzero k-points. Linear bands that go to zero frequency emerge from non-zero k-points. We find that the number of ellipsoids and their positions within the Brillouin zone can be engineered by changing the connectivities of the wire meshes. This gives us a new degree of freedom to tailor the electromagnetic response of metamaterials, leading to unusual wave propagation behaviors.

To illustrate our idea, we start by considering three different types of wire metamaterials as shown in Fig. 1a-c. All of them are composed of metallic wire meshes, which are assumed to be perfect metal in our simulations. The wire meshes are arranged in simple cubic lattices with lattice constants $a$. Corresponding band structures are shown Figs. 1d-f. The band structure in Fig. 1d corresponds to the simple cubic wire mesh shown in Fig. 1a. It has a cutoff frequency of $0.39\,(c/a)$. The existence of a gap from zero frequency up to the cutoff frequency is well known and well understood. If the wires were made up a plasmonic material, the system can be described as a diluted plasmonic system with a renormalized plasmon frequency [1] from an effective medium viewpoint. Let us consider the double wire mesh shown in Fig. 1b, formed by two copies of wire meshes shown in Fig. 1a. As the metallic volume ratio is doubled, one might think that low frequency waves should be more difficult to pass, and hence the low frequency gap should persists. However, the calculated band structure shown in Fig. 1e shows that the double wire mesh has a light cone lying at Brillouin zone center. The group velocity near $\Gamma$ is smaller than the speed of light in vacuum and it is k-dependent due to the anisotropy of the structure. The band structure shown in Fig. 1f corresponds to the double wire mesh structure shown in Fig. 1c. It has a light cone emerging from the zone corner H. We note that the bands in the light cones shown in Fig. 1e and 1f are singly degenerate and they are quasi-longitudinal modes.

The existence of these bands can be understood by solving Maxwell's equations at quasistatic frequencies ($\omega \rightarrow 0$). When $\omega$ approaches to zero, Maxwell's equation reduces to a Poisson's equation $\nabla^2 \varphi = 0$, where $\vec{E} = \nabla \varphi$, with $\varphi$ being the quasistatic potential. According to uniqueness theorem, the solution $\vec{E}$ is unique when the potentials of all metallic wires are well defined. We first consider the case of a single wire mesh shown in Fig. 1a. The radius of wire is $0.1a$. Once the wire mesh's potential $\varphi_1$ is fixed, the system has a unique solution of electric field. One can easily find that the electric field can only be zero in the whole space. Therefore the system in Fig. 1a has no eigen mode at zero frequency and has a cutoff frequency.

Next we come to the double mesh structure of Fig. 1b, where we add another copy of the wire mesh (blue), shifted from the original (red) by a displacement of $(0.3a \ 0.3a \ 0.3a)$. Both wire meshes have the radius of $0.1a$. The two meshes interpenetrate with each other. They have independent quasistatic potentials $\varphi_1$ and $\varphi_2$ since they do not touch each other. The system has a unique $\bar{E}$ when $(\varphi_2 - \varphi_1)$ is given. Thus the system must have one zero frequency mode lying at somewhere (with Bloch k-vector of $(k_x \ k_y \ k_z)$) in the Brillouin zone. Then we discuss the Bloch k of this zero frequency mode. Note that the left end of the red wire has a potential of $\varphi_1$ as the whole wire. By applying Bloch boundary condition, the potential of the right end of this wire should be $\varphi_1 e^{ik_x a}$. Since the left end and the right end are connected by a wire made up of perfect metal, their potential should be equal, i.e., $\varphi_1 = \varphi_1 e^{ik_x a}$. Like, we also have $\varphi_1 = \varphi_1 e^{ik_y a}$ and $\varphi_1 = \varphi_1 e^{ik_z a}$. Hence the zero frequency mode, as well as the light cone of band structure, should lie at the center of Brillouin zone $\Gamma$. Same conclusion can also be arrived by analyzing the blue wire mesh. It can be inferred that the electric fields of this zero frequency mode point from the red wire mesh to the blue wire mesh.

We then consider the wire metamaterial shown in Fig. 1c. The radii of the horizontal thick wire and the two oblique thin wires are $0.1a$ and $0.03a$, respectively. To see the connectivities of the two meshes, we plot a 2x2 supercell in Fig. 2. Similarly, the two meshes interpenetrate with each other and have independent potentials $\varphi_1$ and $\varphi_2$, and the system have one zero frequency mode. Considering the four corners on the upper surface of the unit cell, one has $\varphi_2 = \varphi_1 e^{ik_x a}$, $\varphi_1 = \varphi_2 e^{ik_x a}$, $\varphi_2 = \varphi_1 e^{ik_y a}$ and $\varphi_1 = \varphi_2 e^{ik_y a}$. For the two long thick wires in horizontal direction, we have $\varphi_1 = \varphi_2 e^{ik_z a}$. For the two small metallic patches at the corners of unit cell, we have $\varphi_2 = \varphi_1 e^{ik_z a}$. Then $\varphi_2 = \pm \varphi_1$, i.e., $k = (0 \ \ 0 \ \ 0)$ or $k = (\pi/a \ \ \pi/a \ \ \pi/a)$. Since $\varphi_2 = \varphi_1$ leads to a zero field solution, the zero frequency mode should lie at the Brillouin zone corner H, which is consistent with the band structure in Fig. 1f.

It is notable that in the above analysis about zero frequency mode holds regardless of the detailed geometries of the wires (for example the wires' radii), as long as the two meshes do not contact. Therefore the detailed geometry (like the filling ratio) of the wire meshes will not affect the Bloch k of these zero frequency modes but affect the group velocities near these k-points (Γ or H).

The argument mentioned above can be easily extended to the metamaterials composed of N wire meshes, which have (N-1) zero frequency mode and associated light cones. The locations of these modes are determined by the connectivities of the wire meshes.

Figure 3a shows a metamaterial composed of three metallic wire meshes (red, blue and green). It has a hexagonal lattice with horizontal lattice constant $a$ and vertical lattice constant $d = 0.6a$. Due to the uniqueness theorem, the system has two nonzero and linearly independent $\vec{E}$ solutions at zero frequency. Using the derivation similar to above, one finds that the zero frequency modes lie at $(0 \ \ 0 \ \ -2\pi/3d)$ and $(0 \ \ 0 \ \ 2\pi/3d)$. The computed band structure is shown in Fig. 3c, which shows a light cone emerging between Γ and A as predicted by the quasistatic consideration. The other zero frequency mode can be inferred by applying time-reversal symmetry. Figures 3b and 3d give an example of four wire meshes, which has three zero frequency modes at $(0\ \ 0\ \ \pi/d)$, $(\pi/a\ \ \pi/a\ \ \pi/2d)$ and $(\pi/a\ \ \pi/a\ \ -\pi/2d)$. Similarly, we can design wire metamaterials with index ellipsoid at other k-point of fractional number of reciprocal lattice and can also introduce more index ellipsoids by adding more wire meshes.

In conclusion, we proposed a new type of metallic wire medium composed of several interpenetrating wire meshes. By designing the connectivities of the wire meshes, we can engineer the location of the index ellipsoid in long wavelength limit.

## ACKNOWLEDGMENTS

This work was supported by Research Grants Council, University Grants Committee, Hong Kong (AoE/P-02/12).

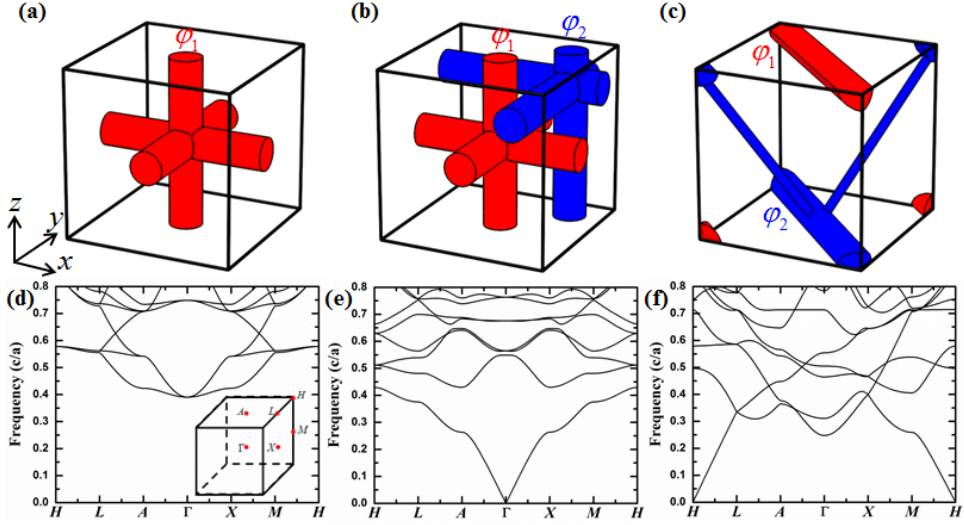

**FIG. 1. Three types of wire mesh metamaterials.** All have a simple cubic lattice with lattice constants of $a$. (**a**) Unit cell composing of a single wire mesh. (**b**) Unit cell containing two wire meshes. Compared to (**a**), another wire mesh (blue) is added, whose node centers at $(0.3a \ \ 0.3a \ \ 0.3a)$. (**c**) Another metamaterial composed of two wire meshes. All the metallic wires (both red and blue) in (**a**)-(**c**) are assumed to be PEC. Their colors just indicate that the two meshes do not contact with each other and should have independent potentials. (**d**)-(**f**) Corresponding band structuresThe single wire mesh structure (**d**) has no eigen mode at zero frequency while the two types of double wire mesh structures have a light cone at Brillouin zone center (**e**) or corner (**f**). The inset in (**d**) shows the Brillouin zone of simple cubic lattice.

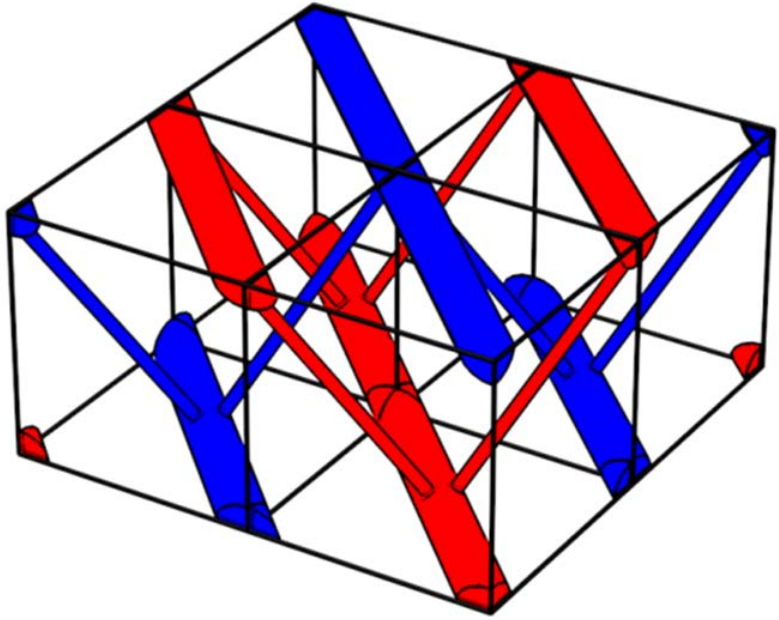

**FIG. 2. 2x2 supercell of the double wire mesh metamaterials in Fig. 1c.** The two wire meshes (red and blue) interpenetrate with each other and do not contact.

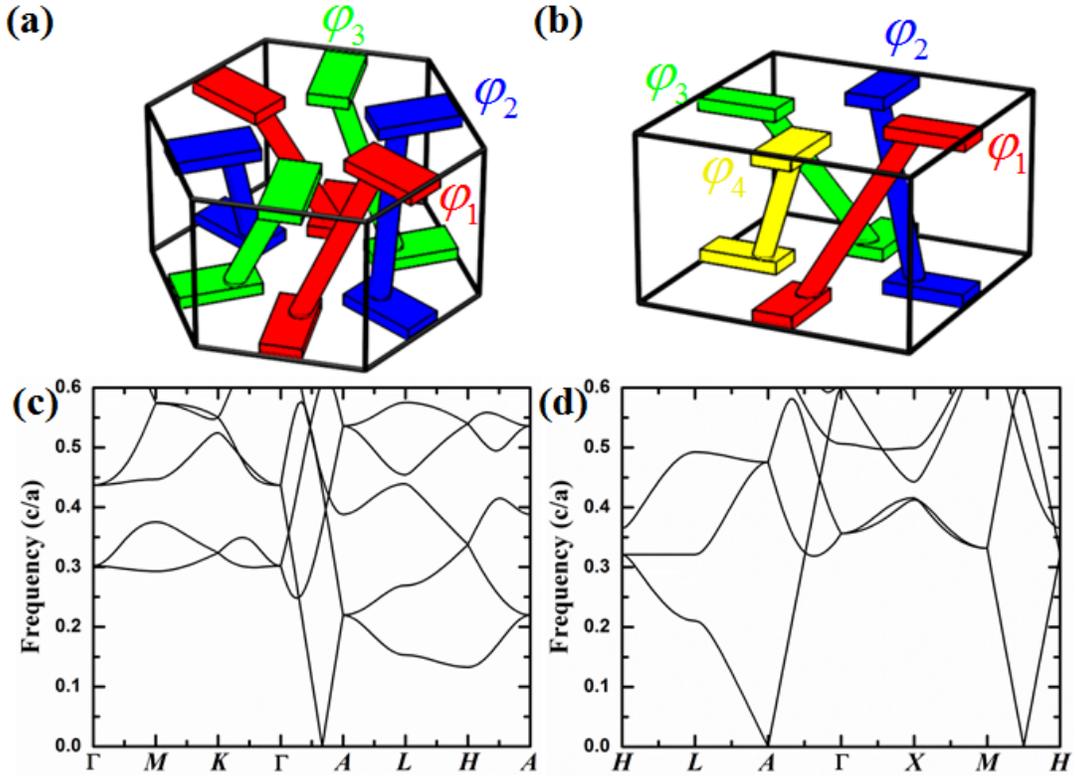

**FIG. 3. Wire metamaterials composed of three or four wire meshes.** (a) Unit cell of the wire metamaterial composed of three meshes. (b) Unit cell of the wire metamaterial composed of four meshes. (c)/(d) Corresponding band structures. Note that the linear bands can emerge from arbitrary k-points inside the Brillouin zone.